\documentclass[mypaper,7pt,twoside]{CoAst}
\usepackage{epsf,graphicx,fancyhdr}
\renewcommand{\chaptermark}[1]{\markboth{#1}{}}

\newcommand{\kopf}{\small\itshape Comm. in Asteroseismology\\ Vol. number, publication date (will be inserted in the production process)}
\newcommand{\Authors}[1]{\begin{center}\normalsize\bf\sf #1 \end{center}}

\renewcommand{\author}[1]{\begin{center}\normalsize\bf\sf #1 \end{center}}
\newcommand{\Address}[1]{\begin{center}\small\sf #1 \end{center}}

\renewenvironment{abstract}{\section*{Abstract}\normalsize\sf}{}
\newcommand{\References}[1]{\begin{flushleft}{\large References\\}\vspace*{2mm}\small #1 \end{flushleft}}

\newcommand{\chapterDSSN}[2]{\chapter[\sf\normalsize #1\\ \footnotesize \hspace*{5mm}by #2 \sf\normalsize][]{#1\\}\rhead[\fancyplain{}{\sf\footnotesize \center{#1}}]{\fancyplain{}{\sffamily\thepage}}\lhead[\fancyplain{\kopf}{\sffamily\thepage}]{\fancyplain{\kopf}{\sf\footnotesize \center{#2}}}}

\renewcommand{\chaptername}{}
\newcommand{\acknowledgments}[1]{\vspace*{5mm}\noindent\begin{bf}Acknowledgments. \end{bf} #1}

\begin{document}
\sf

\def\uv{uvby-\beta \ }

\chapterDSSN{On the nature of the  $\delta$ Scuti
star star HD 115520}{L. Fox Machado, J. H. Pe\~na, G. Mu\~noz and B.
Vargas}

\Authors{L. Fox Machado$^1$, J. H. Pe\~na$^2$, G. Mu\~noz$^3$, B.
Vargas$^3$} \Address{$^1$ Observatorio Astron\'omico Nacional,
Instituto de Astronom\'{\i}a,
Universidad Nacional Aut\'onoma de M\'exico, Ensenada B.C., Apdo.
Postal 877,
M\'exico\\
$^2$ Instituto de Astronom\'{\i}a, Universidad Nacional Aut\'onoma
de M\'exico, M\'exico D.F., Apdo. Postal 70-264,
M\'exico\\
$^3$ Escuela Superior de Ingeniería Mec\'anica y El\'ectrica,
Instituto Polit\'ecnico Nacional, Av. IPN s/n, 07738 M\'exico, D.F.,
M\'exico}

\noindent
\begin{abstract}
As a continuation of the study of the newly found $\delta$ Scuti
star HD 115520, we present a period analysis of  recently acquired
photometric data covering four nights, as well as some conclusions
on the nature of this star.
\end{abstract}

\section{Introduction}

In 2007 Pe\~na et al. (2007, Paper I) confirmed  the belonging of HD
115520 to the $\delta$ Scuti class which was considered as a
standard star in a 2005 observing run. From the relatively large
scatter shown, Pe\~na et al.(2006) consider it as a variable
candidate. With this in mind, new data were acquired in two new
nights in 2006 which established it as a $\delta$ Scuti star. In the
present paper we present new observations which were performed in
2007 with the same instrumentation over a period of four nights and
which have served to determine its periodic content. The found
frequencies explain the behavior of both seasons separated by more
than one year.

\section{Observations} These were taken at the Observatorio Astr\'onomico
Nacional, M\'exico using the 1.5 m  telescope to which a
spectrophotometer was attached. The observing season was carried out
on four consecutive nights in March and April, 2007. The following
observing routine was employed: a multiple series of integrations
was carried out, consisting of five 10 s integrations of the star to
which one 10 s integration of the sky was subtracted. Two reference
stars were also observed C1: HD116879 and C2: HD114311. These were
observed in the following sequence to optimize the time coverage of
the variable: V, sky, C1, V, V, C2, V. A series of standard stars
was also observed at the beginning and at the end of each night to
transform the data into the standard system. The absolute
photometric values of the 2007 campaign are provided in an archive.
The accuracy of the season is deduced from the differences between
the reduced and the previously reported values of the standard
stars. Due to the fact that the last night was of lower quality, and
hence less accurate, the mean values of the differences are
calculated only from the standards of the first three nights. They
are: 0.015, 0.008, 0.007, 0.011 mag for V, $(b-y)$, $m_1$, and
$c_1$, respectively.

However, since the amplitude of the star is typical 
of a  $\delta$ Scuti star ($\sim$ 20 mmag, see 
Figure 4), we preferred to analyze the data
for the periodic content through differential photometry in the $y$
filter for which use was made of the reference  stars C1 and C2 to
increase the accuracy of the photometry to thousands of magnitude.
Table 1 lists the characteristics of the observed stars. A magnitude
value of the reference stars was interpolated at the time of the
variable and the final values, to which the average value of each
night was subtracted, are presented in Table 1. The whole reduction
procedure is shown in Figure 1 for the night of March 30/31. The
2006 season was reduced in the same fashion to match the newly
acquired data.

\begin{table*}[!t]\centering
  \setlength{\tabcolsep}{1.0\tabcolsep}
 \caption{ Characteristics of the observed stars. The spectral types were taken from the SIMBAD database.}
  \begin{tabular}{lccccccccc}
\hline\hline
 $ID$ & RA & Dec & V & $(b-y)$ &  $m_1$ &  $c_1$ & N & SpTyp \\
\hline
HD 115520& 13 17 21.4 &+30 36 45.5 & 8.435 & 0.132 & 0.171 & 0.806  & 459 &F0  \\
HD 116879& 13 26 06.9 &+30 42 08.2 & 7.953 & 0.272 &0.144  & 0.634 & 120 & F5  \\
HD 114311& 13 09 28.9 &+30 26 06.3  & 9.037 & 0.334 & 0.157 & 0.474  & 122 &F6V \\
\hline
\end{tabular}
\end{table*}

\begin{figure}[t]
\includegraphics[width=12cm]{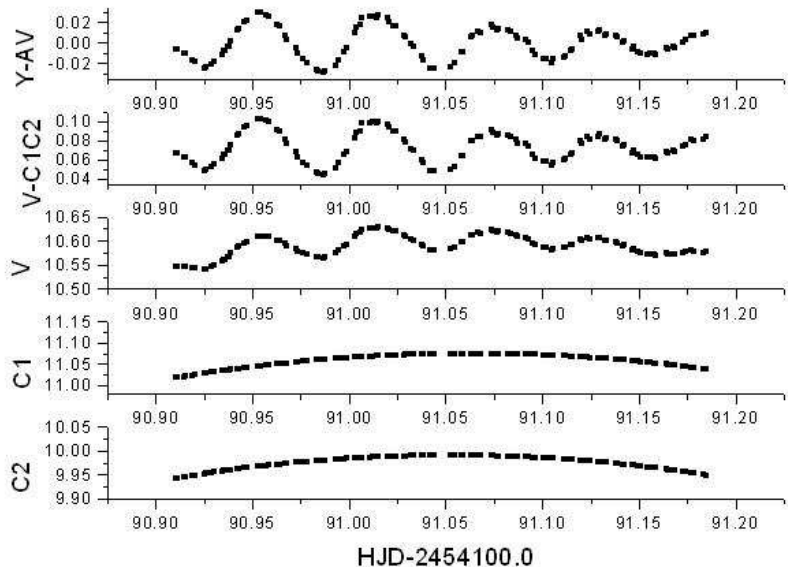}
\caption{Y variation of the observed stars HD 115520, C1 and C2 on
the night of March 30/31, 2007 . Y axis is V in magnitudes, X axis
is time (time shown=HJD - 2454100.0)}
\end{figure}

\section{Frequency determination}
With the relatively few data points acquired in the 2006 season
(only two short nights) we were able to demonstrate the star's
variability and found evidence of at least two close frequencies
which might explain the resulting beating behavior of the light
curve. Since the new photometric data is constituted of four long
consecutive nights, we are now able to determine  the pulsational
frequencies with greater precision. Two numerical packages were
utilized: Period04 (Lenz and Breger, 2004) and ISWF (Alvarez et al.,
1998). With Period04 the first run examined gave a frequency of
17.8643 c/d with an amplitude of 0.0140 mag in the frequency
interval between 0 and 30 c/d with a step rate of 0.0150.
Prewhitening of this frequency consecutively yielded  the results
shown in Figure 2. On the other hand, the ISWF package yielded the
following frequencies (in c/d) listed in diminishing amplitudes (in
parentheses, in mmag) 17.850 (13.877); 14.7786 (10.334); 17.4527
(6.415); 13.5217 (4.236) and 18.1831 (3.973).

\begin{figure}[t]
\begin{center}
\includegraphics[width=10cm]{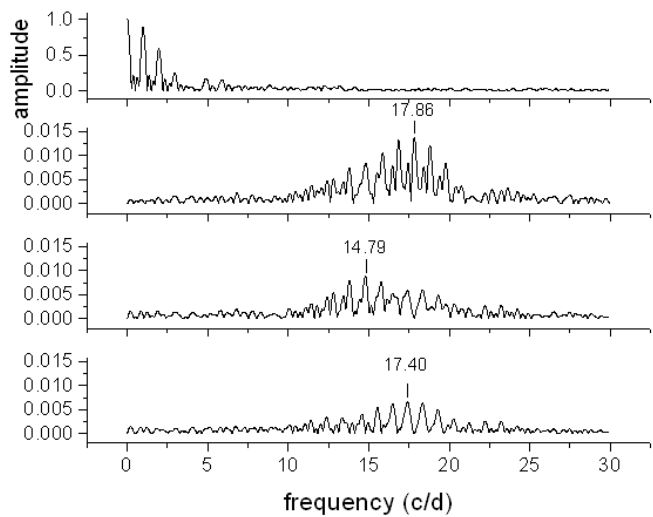}
\caption{Periodograms of the four consecutively observed nights in
2007. From top to bottom, window, first frequency obtained at
17.86c/d, periodogram after prewhitening this with a resulting peak
at 14.79c/d, and finally the prewhitened histogram of the two
previously determined frequencies with a peak at 17.400c/d.}
\end{center}
\end{figure}

As can be seen, the two previously determined main frequencies,
although slightly numerically different are confirmed. In the 2006
season we obtained 18.82 and 14.63 c/d. Given the complex window
function of observations on only two nights from only one
observatory, we might consider them the same. On the other hand,
when the whole dataset was utilized with a step rate of 0.00015,
Period04 yielded peaks at 17.8373 and 14.7537 c/d, (see Figure 3 and Table 2). 
The rest of the frequencies might be disregarded because they do not
significantly improve the residuals. Their  peaks are
indistinguishable from each other due to the aliasing caused by the window function. Therefore, we will consider as definitive only the first two frequencies 
listed in Table 2. Figure 4 shows the light curves of the six observed nights.

\begin{figure}[t]
\begin{center}
\includegraphics[width=10cm]{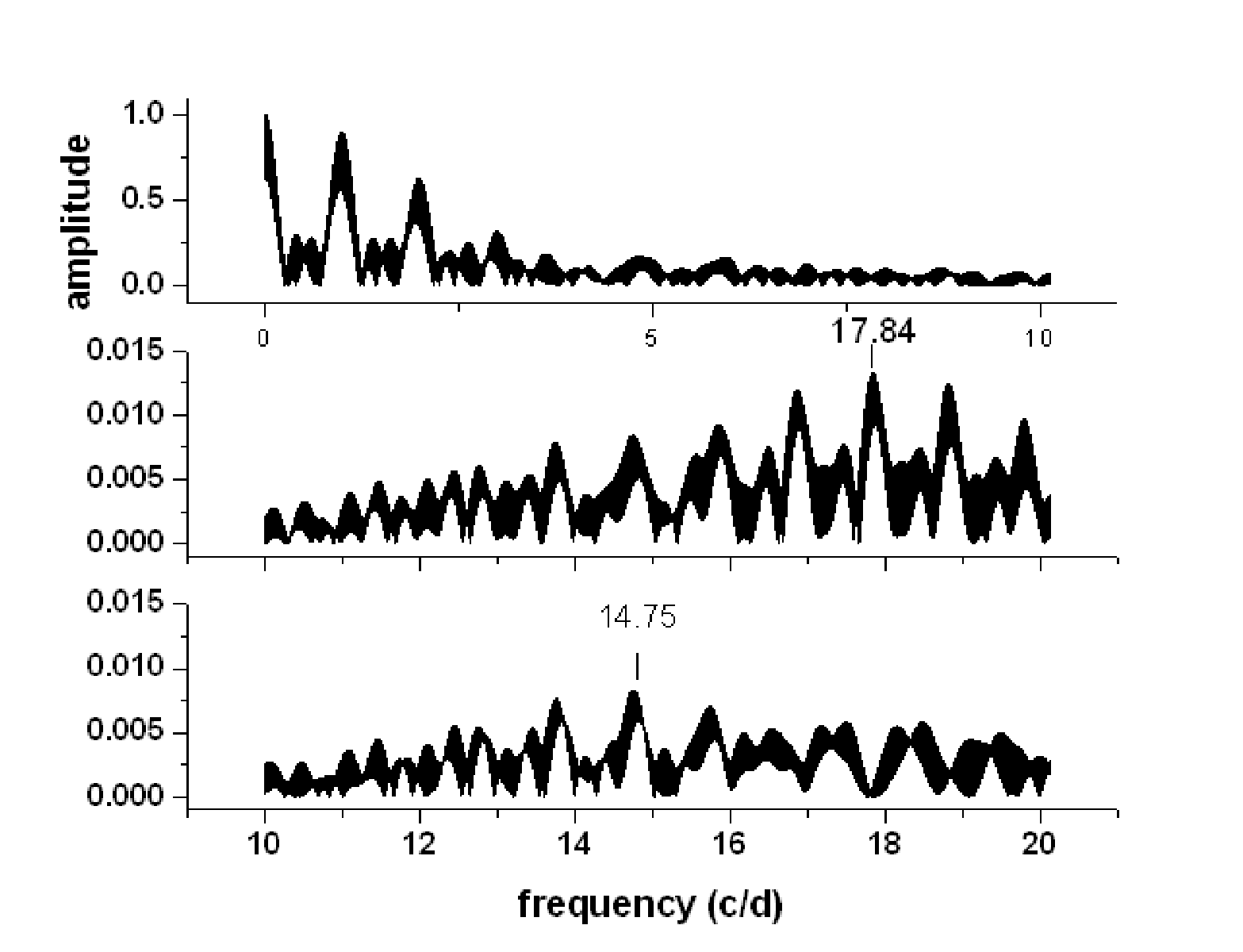}
\caption{Periodograms of all the observed nights. From top to
bottom, window, first frequency obtained at 17.8375c/d, periodogram
after prewhitening this with a resulting peak at 14.7537c/d, and
finally the prewhitened histogram of the two previously determined
frequencies with a peak at 16.5121c/d.}
\end{center}
\end{figure}

\begin{table*}[!t]\centering
  \setlength{\tabcolsep}{1.0\tabcolsep}
 \caption{ Frequencies, amplitudes and phases derived }
  \begin{tabular}{llllll}
\hline\hline
 & Frequency (c/d)    &    Amplitude (mag) & Phase    \\
\hline
F1 & 17.8375  &  0.0131 &  0.1028\\
F2 & 14.7537  &  0.0108 &  0.2612\\
F3 & 16.5121  &  0.0070 &  0.5646\\
\hline
\end{tabular}
\end{table*}

\begin{figure}[t]
\begin{center}
\includegraphics[width=10cm]{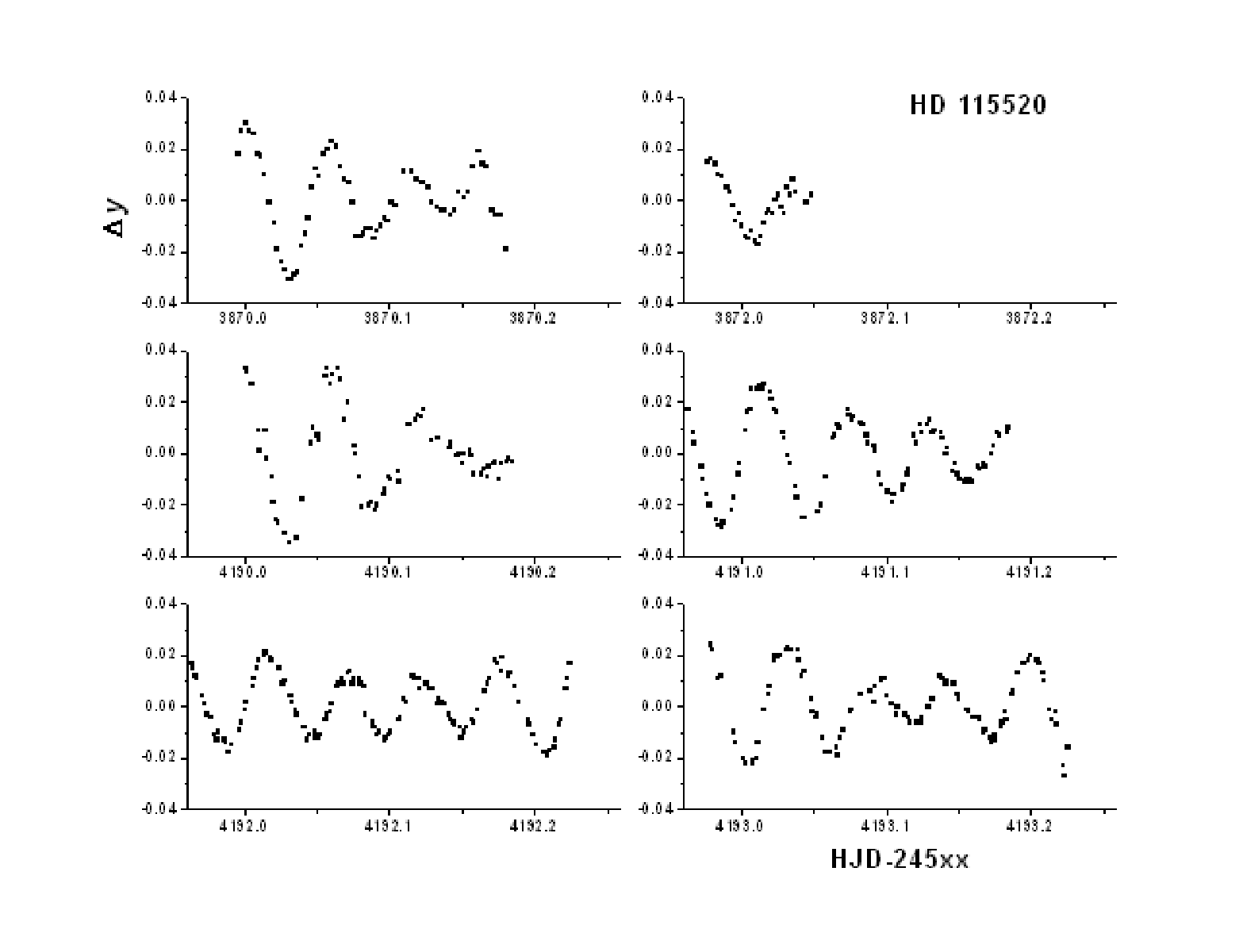}
\caption{$y$ variation of HD 115520 (dots). Y axis is y in
magnitudes, X axis is time. }
\end{center}
\end{figure}

\section{Physical parameters}
As it has been already described in Paper I, we carried out a
well-known procedure to determine reddening as well as unreddened
colors using the photometric mean $\uv$ values reported in Table 3.
Table 4 lists the reddening, the unreddened indexes, the absolute
magnitude, and the distance. Its position on the $[m_1]-[c_1]$
diagram established it to be an A8V star. Its temperature and log of
surface gravity can be determined by locating HD 115520 in the
$(b-y)_0$ vs. $c_0$ grids of Lester et al. (LGK86) (Figure 5); the
values we determine are 7700 K and 4, respectively. As was stated in
Paper I, we compared our results with those in a paper by Behr
(2003) who found  an effective temperature $T_{\rm eff}$ of 8199
(+449,-317), a log g 4.63 (+0.34,-0.23), an [Fe/H] 0.62 (+-0.13) and
a stellar type belonging to the main-sequence for this star.
Although Behr (2003) has evaluated physical parameters for this
star, and his numerical values coincide with ours, we feel that we
have more data to determine the physical characteristics.
Nevertheless, we have employed his reported metallicity of HD 115520
to discriminate between the models that explain the star's behavior.

\begin{table*}[!t]\centering
  \setlength{\tabcolsep}{1.0\tabcolsep}
 \caption{ Mean values of the $\uv$ photometry of HD 115520 from the two seasons }
  \begin{tabular}{lllllllll}
\hline\hline
 & average &   sigma     &    N     \\
\hline
V (mag)      & $8.4305$ & $0.0178$ & $579$ \\
$(b-y)$ (mag) & $0.1334$ & $0.0070$ & $584$ \\
$m_1$   & $0.1701$ & $0.0051$ & $580$ \\
$c_1$   & $0.8068$ & $0.0139$ & $584$\\
$\beta$ & $2.8108$ & $0.0133$ & $67$ \\
\hline
\end{tabular}
\end{table*}

\section{The evolutionary status of HD 115520}

The determination of the evolutionary stage of a field star requires
precise estimates of its global parameters. In the case of HD 115520
the distance as determined from Str\"omgren photometry is 140 pc
which leads an $M_V$ of 2.86 mag by using the calibrations of Shobbrook (1984).
 On the other hand, the distance value of
300 pc estimated from a parallax of 3.29 $\pm$ 0.97 mas  provided by  the Hipparcos catalogue 
(Perryman et al. 1997) yields  an $M_V$ of 1.02 mag
which is quite different from the photometric one. This ambiguity
can be explained by the uncertainties in the determination of each
measured distance. The large relative error ($\sigma (\pi))/ \pi \sim
0.30$ ) of the Hipparcos parallax for HD 115520 implies an $\sigma
(M_V)  > 0.5$ mag, whereas in the present paper the uncertainty in
the apparent  magnitude  derived as explained in Pe\~na \& Sareyan (2006)
from the standard deviation of 579 data points of the two seasons gives 
an $m_V = 8.4305 \pm 0.0178$ (see Table 3) and an $\sigma (M_V)  < 0.1$ mag.
Although this latter value  does not include  
the uncertainty in $M_V$ due to the photometric calibrations which can be 
as large as 0.3 mag for early type stars (e.g. Balona \& Shobbrook 1984), 
we think that the photometric distance is more reliable than the trigonometric
one because different photometric calibrations (Balona \& Shobbrook (1984) and Nissen (1988)) lead to similar distance values for HD 115520.
Furthermore,
similar values of $m_V$ for HD 115520 have already been
reported in previous papers (Olsen 1983, Crawford \& Perry 1989,
Paper I). Therefore,  we  will use the photometrically 
determined distance
to tray to establish the evolutionary status of HD 115520.

Figure 6 shows the observed position of HD 115520 (asterisk) in the
HR diagram and its associated uncertainty (cross upon the asterisk).
PMS and post-MS evolutionary tracks giving a range of masses between
1.45-1.60 $M_{\odot}$ for HD 115520 are shown with dotted and
continuous lines respectively. These evolutionary sequences were
computed by using the CESAM evolution code (Morel 1997) with an
input physics appropriate to $\delta$ Scuti stars and a chemical
initial composition of $Z=0.013$ and $Y=0.28$. Also shown are the
theoretical pre-MS instability strip boundaries of the first three
radial modes obtained by Marconi \& Palla (1998).

According to the models depicted in Fig. 6  HD 115520 could either
be  in pre-MS stage with an age between 15-20 Myr or post-MS stag
with an age between 500-700 Myr. In the former case,  the age was
estimated as the time spent by the star travelling from the
birthline to the ZAMS in the HR diagram according to the isochrones
given by Tout et al. (1999).

As shown by  Suran et al. (2001) non-radial oscillation spectra in
the low frequency domain can be used to discriminate  between the
pre- and post-MS stage. In the present case, however, this is seldom
possible since the two detected peaks in HD 115520 are most likely
due to radial oscillations. In fact, we have tried to reproduce the
observed periods computing linear adiabatic pulsation models of HD
115520 for some selected pre- and post-MS models located within the
error box  in Figure 6, but no satisfactory fit between observed and
theoretical frequencies was found. Therefore,  more observational
efforts are required to establish the true nature of this
interesting object.

\begin{table*}[!t]\centering
  \setlength{\tabcolsep}{1.0\tabcolsep}
 \caption{Reddening and unreddened parameters of HD 115520 }
  \begin{tabular}{lccccccccc}
\hline\hline
 $E(b-y)$ & $(b-y)_0$ & $m_0$ & $c_0$ & $V_0$ &  $M_v$ &  $DM $ & dst (pc) \\
\hline
0.000& 0.135 &0.170 & 0.807 & 8.43 & 2.68 & 5.75 & 141 \\
\hline
\end{tabular}
\end{table*}

\begin{figure}[t]
\begin{center}
\includegraphics[width=10cm]{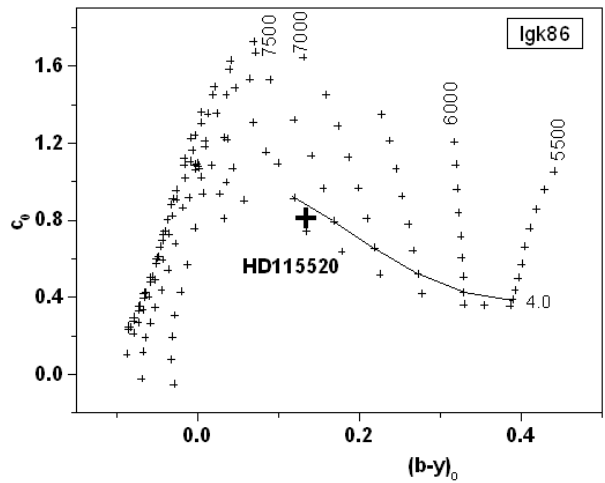}
\caption{Location of the photometric data of HD 115520 in the grids
of LGK86.}
\end{center}
\end{figure}

\begin{figure}[t]
\begin{center}
\includegraphics[width=10cm]{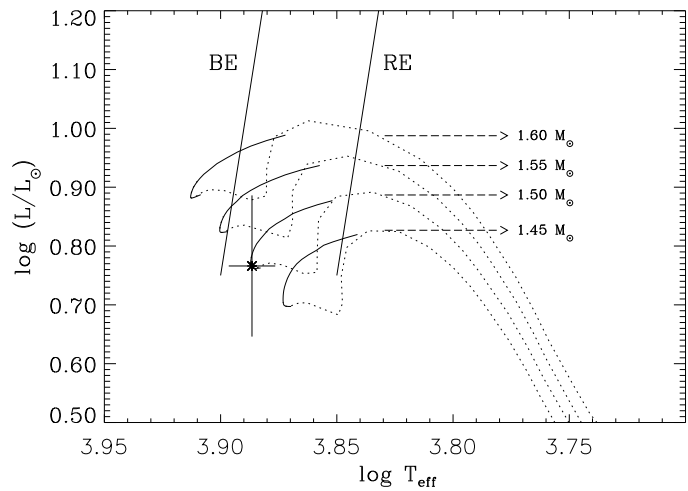}
\caption{Position of HD 115520 in the HR diagram.}
\end{center}
\end{figure}

\section{Conclusions}
We have presented the analysis of new $uvby$ photometric
observations of $\delta$ Scuti star HD 115520 carried out during
four nights in March and April, 2007 at the Observatorio
Astron\'omico Nacional, M\'exico. These data were added to the
previously observed two nights in 2006 resulting
 a total of 580 data points of $uvby$ photometry which allowed us to
 search for the true nature of this $\delta$ Scuti variable. The two
oscillations frequencies detected in 2006 have been confirmed in
this season. We have found that both  stages pre-MS and post-MS are
possible to account for the observed luminosity and temperature of
the star. We thus conclude that HD 115520 represents an good
candidate for asteroseismological studies of young $\delta$ Scuti
stars.

\acknowledgments{We would like thank the assistance of the staff of
the OAN during the observations. This paper was partially supported
by Papiit IN108106. GM and BV thank the OAN for allowing the use of
the telescope time and to Dr. Jorge Sosa of IPN for the support. J.
Miller and J. Orta did the proofreading and typing, respectively.
This article has made use of the SIMBAD database operated at CDS,
Strasbourg, France and ADS, NASA Astrophysics Data Systems hosted by
Harvard-Smithsonian Center for Astrophysics. }

\References{
Alvarez, M.,  et al., 1998, A\&A 340, 149\\
Balona, L.A., \& Shobbrook, R.R., 1984, MNRAS 211, 375
Behr, B. B., 2003, ApJS 149, 101\\
Breger, M., 1979, PASP 91, 5\\
Lenz, P. \& Breger, M., 2005, CoAst 146, 53\\
Lester, J. B., Gray, R. O. \& Kurucz, R. I., 1986, ApJ 61,
    509\\
Marconi, M., \& Palla, F., 1998, ApJ 507, L141\\
Morel, P., 1997, A\&AS 124, 597\\
Nissen, P., 1988, A\&A 199, 146\\
Pe\~na, J. H., and Sareyan, J. P., 2006, RevMexAA 42, 179\\
Pe\~na, J. H., Sareyan, J. P., Cervantes Sodi, B. et al., 2007, RevMexAA 43, 217 (Paper I)\\
Olsen, E. H., 1983, A\&AS 54, 55\\
Perryman, M.A.C, et al., 1997, A\&A 323, L49\\
Rodriguez, E. \& Breger, M., 2001, A\&A 366, 178\\
Shobbrook, R. R., 1984, MNRAS 211, 659\\
Suran, M., et al., 2001, A\&A 372, 233\\
Tout, C.A, et al., 1999, MNRAS 310, 360\\

 }

\end{document}